\documentstyle[preprint,prl,aps]{revtex}
\tightenlines
\begin{document}
\preprint{\it submitted to Physical Review Letters}
\title{The local spectrum of a superconductor as a probe of interactions
between magnetic impurities}
\author{Michael E. Flatt\'e and David E. Reynolds}
\address{Department of Physics and Astronomy, University of
Iowa, Iowa City, Iowa 52242}
\maketitle
\begin{abstract}
Qualitative differences in the spectrum of a superconductor near 
magnetic impurity pairs with moments aligned parallel and
antiparallel are derived. A proposal is made for a
new, nonmagnetic scanning tunneling spectroscopy of magnetic
impurity interactions based on these differences. Near parallel impurity pairs
the mid-gap localized spin-polarized
states associated with each impurity hybridize and 
form bonding and anti-bonding molecular states with different energies. 
For antiparallel impurity moments the states do not
hybridize; they are degenerate.
\end{abstract}
\vfill\eject

The relative orientation of the moments of two magnetic 
impurities embedded nearby in a metallic nonmagnetic host will 
depend on the significance of
several electronic correlation effects, such as direct exchange, double
exchange, superexchange, and RKKY. Each of these effects produces 
characteristic moment orientation; the RKKY interactions
can align moments either parallel or antiparallel depending on the 
impurity separation. Reliable experimental measurements of the moment 
orientation as a function of impurity separation could identify 
the origin of magnetism
in alloys of technological significance, such as the metallic ferromagnetic
semiconductor GaMnAs\cite{Bers} which may eventually play a 
crucial role in semiconductor-based
magnetoelectronics\cite{Ohno}. Such measurements should also
clarify the interplay between
metallic and magnetic behavior in layered oxides, such as the
high-temperature superconductors.
In this work we propose, based on theoretical calculations,
a robust experimental technique for the systematic and 
unambiguous experimental determination of moment
alignment as a function of impurity separation. 

We demonstrate that in an electronic system with a gap
there is a fundamental difference between the electronic states
localized around parallel and antiparallel impurity moments.
Around parallel impurity moments there are
mid-gap molecular states (similar to bonding and antibonding
states in a diatomic molecule).
Around antiparallel impurity
moments the states remain more atomic-like and are
degenerate. 
This qualitative difference in the
spectrum of an impurity pair provides a robust technique of
determining the impurity-impurity interaction via {\it nonmagnetic}
scanning tunneling spectroscopy (STS). 
The essential condition for practical application of this
technique will be whether the splitting of the states around
parallel impurity moments is large enough to be observed
spectroscopically.

The gapped system we consider in detail
is the superconductor NbSe$_2$, which
is chosen for its extremely favorable surface properties for 
STS and for its quasi-two-dimensional electronic structure. 
STS has already been 
used to examine 
the localized states which form near isolated magnetic impurities
on the surface of superconducting niobium\cite{Eigler,Shiba}.
We have calculated the energies and spatial structure
of the electronic states near impurity pairs in NbSe$_2$
essentially exactly within mean-field theory.
These calculations indicate that the size of the splitting of states
around parallel impurity moments in NbSe$_2$ is measurable --- they
are split by a sizable percentage of the energy gap even for impurity
moment separations of order 30\AA.

A nonmagnetic spectroscopy of magnetic impurity interactions is 
also plausible in
a much wider range of materials. 
The localized spin-polarized 
states upon which the technique is based 
occur near magnetic impurities 
in most systems where there is a gap in
the single-particle density of states at the chemical potential, whether or
not the gap originates from superconductivity.
Even when there is no true gap, if the density of states is substantially
reduced at the chemical potential sharp resonances similar to the
localized states will form (this has been predicted and recently observed for
$d$-wave
superconductors\cite{Balatsky,Flatte,Davis}). Resonances around parallel and
antiparallel impurity pairs
show similar qualitative features to localized states.

If the energy scales of moment formation and interaction are
much greater than those responsible for creating the gap it is also possible
to infer the impurity interaction within a material in its high-temperature
metallic phase from spectroscopic
measurements on the same material 
in a low-temperature superconducting phase.
In this the STS procedure is similar to traditional
``superconducting spectroscopy'',\cite{Maple} where the dependence on
impurity concentration of
the superconducting transition temperature 
$T_c$ or the specific heat discontinuity at $T_c$ 
is used to determine the
presence and rough magnitude of a single impurity moment. However, whereas 
single-impurity information can often be extracted from such 
measurements in the dilute limit, pairwise impurity 
interactions are much more difficult to infer
from macroscopic properties like $T_c$ which
depend on an ensemble of local configurations. 

We note that the technique described here is remarkably
non-invasive compared to alternate methods. 
The use of a magnetic tip to probe the magnetic properties
of a sample\cite{Wiesendanger} may 
distort the natural surface orientation of moments. An alternative
nonmagnetic STS technique that has been proposed, which involves
a superconducting tip\cite{Davis2}
in a Tedrow-Meservey geometry\cite{TM}, requires either an external or 
surface-induced magnetic field to spin-split the superconducting DOS of the tip.
Finally, the
use of spin-polarized tunneling from a GaAs tip relies on a fixed
orientation of the magnetic structure on the surface 
relative to that of the optically generated spin-polarized population 
in the tip\cite{GaAs}.

To understand the origin of the non-degeneracy of states around
parallel moments and the degeneracy of states around
antiparallel moments 
consider a
heuristic picture of the two-impurity system in an
isotropic-gap superconductor. For parallel alignment of
the impurity moments 
only quasiparticles of one spin direction (assumed to be spin up)
will be attracted to the impurity pair. Any localized state will thus be spin
up.  If the two impurities are close
their two spin-up atomic-like states will hybridize and 
split into molecular
states just as atomic levels are split into bonding and
antibonding states in a diatomic molecule. Thus there will
be two non-degenerate states apparent in the spectrum. This is shown
schematically in the top section of Fig. 1, where the potential for spin up
quasiparticles is shown on the left (Fig. 1A) and for spin down
quasiparticles is shown on the right (Fig. 1B). The
potential for spin-down quasiparticles is everywhere
repulsive, so no spin-down localized states will form.

The situation for antiparallel aligned spins, shown
on the bottom of Fig. 1, is
quite different. The effect of the second impurity on the
state around the first is {\it repulsive} and so does not
change the state energy much unless the impurities are very
close. Furthermore the Hamiltonian has a new symmetry in this case: it
is unchanged under the operation which 
both flips the quasiparticle spin and inverts space
through the point midway between the two impurities. This
operation changes the potential of Fig. 1C into that of
Fig. 1D. Thus
instead of split states we find two degenerate atomic-like states of
opposite spin, localized around each of the two impurities.

Detailed results for NbSe$_2$ are obtained by
solving the following lattice-site mean-field Hamiltonian
self-consistently:
\begin{equation}
H = -\sum_{\langle ij\rangle,\sigma}
t_{ij}c_{i\sigma}^\dagger c_{
j\sigma} + \sum_i\left[\Delta_{i}
c_{i\uparrow}^\dagger c_{i\downarrow}^\dagger+\Delta_{i}^*
c_{i\downarrow}c_{i\uparrow}\right] +
V_{S1}(c_{1\uparrow}^\dagger
c_{1\uparrow} - c_{1\downarrow}^\dagger c_{1\downarrow})
+ V_{S2}(c_{2\uparrow}^\dagger
c_{2\uparrow} - c_{2\downarrow}^\dagger c_{2\downarrow}),
\end{equation}
where $c^\dagger_{i\sigma}$ and $c_{i\sigma}$ create and
annihilate an electron at lattice site $i$ with spin $\sigma$.
The impurities reside at lattice sites $1$ and $2$, 
the $t_{ij}$ are the hopping matrix elements and
the $\Delta_{i}$ are the values of the superconducting order parameter.
NbSe$_2$ has a triangular lattice, and the normal-state band
structure can be modeled with an
on-site energy of $-0.1$~eV and with nearest-neighbor and 
next-nearest-neighbor hopping matrix elements of
$-0.125$~eV. These are determined from a tight-binding fit\cite{param} to 
{\it ab initio} calculations of the electronic structure\cite{NbSe2}. 
The superconducting pairing interaction is modeled with an on-site attractive
potential which yields the experimental order parameter
$\Delta = 1$~meV. 
The inhomogeneous order parameter $\Delta_i$ is determined self-consistently
from the distorted electronic structure in the vicinity of
the impurities. 
We consider equivalent parallel 
($V_{S1} = V_{S2}$) or antiparallel ($V_{S1} = -V_{S2}$) impurity moments.

This model assumes the impurity spins behave as classical
spin (see Refs.~\cite{Eigler,Shiba,Flatte}). Classical spin
behavior has been seen, for example, for Mn and Gd
impurities on the surface of niobium\cite{Eigler}.
The electronic structure in this model, 
including quasiparticle state energies and spatial structure, can be found 
rapidly and accurately by inverting the Gor'kov equation in a
restricted real-space region including the two impurities,
as described in Ref.~\cite{Flatte}. 
Measurements of the spatial structure of these states and of the values of the
splitting between states can serve as a sensitive test of the model of the
electronic structure of this material and of the impurity potential for
a given atom.

Figure 2A shows the energies of the localized states in NbSe$_2$ for
parallel spins (red) and antiparallel spins (black) for a
sequence of impurity spacings which are multiples of the in-plane
nearest-neighbor vector of the NbSe$_2$ lattice.
The splitting of the bonding and antibonding states
oscillates over a distance scale comparable to 
the Fermi wavelength
of NbSe$_2$ along this direction. The splitting is
proportional to the probability of a quasiparticle at one
impurity propagating to the other, which is a measure of the
coupling of the two atomic-like states. 
At large distances
state energies for parallel and antiparallel moments approach
the single impurity state energy, indicated on the right side of
Fig. 2A. 
Figure 2BC shows the spatially integrated
change in density of states due to the impurity pair
for these impurity separations. The density of states (DOS)
of a quasiparticle of energy $E$ in a superconductor has an electron 
component at energy $E$ and a hole component at energy $-E$, so a 
single state will produce two peaks in the DOS unless it is closer to 
$E=0$ than the linewidth. 
That
linewidth is determined by thermal broadening in the
metallic probe tip, which for these 
plots is assumed to be $0.05$~meV$ = 0.6$K.
The gap in the homogeneous DOS extends from $-1$~meV to $1$~meV in NbSe$_2$,
so the variation in state energies is a substantial fraction of this gap.
The clear distinction between parallel and antiparallel impurity 
moments in the DOS is only limited by the linewidth of the states. 

A tunneling measurement of the DOS using a broad-area contact would yield the
spectrum of an ensemble of impurity separations, hence STS (which 
measures the local DOS, or LDOS) is the
ideal method for examining a single configuration of impurities.
Before describing the distinct spatial differences in LDOS
measurements between parallel and antiparallel alignments of
impurity pairs we show the single impurity result
in Fig. 3. The spatial structure of the electron and hole components of the
LDOS are independently measurable by STS and can be quite different in detail.
In this work we will show only the spatial structure of
the hole component --- similar gross 
structure is seen in the electron-like LDOS. 
Figure 3 shows
the six-fold symmetric LDOS for NbSe$_2$  for 
$V_S = 200$~meV at an energy of $-0.19$~meV. The units
are Angstroms and the nearest-neighbor spacing is 3.47\AA. 

The details of the 
spatial structure can be traced directly to the normal-state
electronic structure of NbSe$_2$.\cite{Flatte} We note that
the local hopping matrix elements and the local nonmagnetic
potential will differ near the impurity atoms. We find that
moderate changes in these quantities do not significantly
change the magnitude of the splitting of the even and odd
parity states. This relative insensitivity occurs because
the splitting is largely dependent on the amplitude for a
quasiparticle to propagate from one impurity site to the
other. Careful comparison of a measured LDOS and Fig.~3
would allow the determination of the changes in the local
hopping and the nonmagnetic potential for this case.

Plots of the LDOS for two impurities in NbSe$_2$ 
separated by four lattice spacings (13.88\AA) are
shown in Fig.~4A-D. They demonstrate via their spatial structure the
qualitative differences among 
different types of molecular states possible 
around an impurity pair.
Figure 4A is the bonding state (energy $-0.10$~meV)
and Fig. 4B shows the
antibonding state ($-0.26$~meV). 
The impurities are at the same 
sites in each of Fig.~4A-D, labeled $1$ and $2$ in Fig.~4B.
As expected from the symmetry of 
these states, the antibonding state has a
nodal line along the mirror plane (indicated in red)
between the two impurities. No such nodal line occurs in Fig.~4A
--- in contrast the state is enhanced along the nodes.

The nonmagnetic STS probe cannot resolve the spin direction of the electronic
states around the impurities, so around antiparallel impurity moments it
detects both states. The sum of the LDOS for the two atomic-like states is
symmetric around the mirror plane.
Figure 4C is the LDOS at the energy for the two degenerate
states around antiparallel impurity spins ($-0.28$~meV). 
The states are much more
diffuse than the bonding state in Fig.~4A due to the repulsive nature 
of one impurity. Figure 4D shows the experimentally inaccessible
spin-resolved LDOS, showing the LDOS of holes with the spin direction
attracted to the impurity on the left. The spin-resolved LDOS at the 
impurity on the left is two orders of magnitude
greater than at the impurity on the right. Thus the individual localized
states are quite atomic-like.

We have assumed throughout that the impurity moments are locked either
parallel or antiparallel. If the alignment is intermediate between the two
cases then the spectrum shows non-degenerate states split less than in 
the parallel case. If there is some flipping of moments between parallel and
antiparallel alignment on a
timescale longer than the time required for the quasiparticle states
to realign with the moments then the spectrum would be a
linear superposition of the antiparallel and parallel spectra. If this is an 
activated process, this energy of activation of moment
flipping could be
easily distinguished by examining the temperature
dependence of the spectrum. 

This work describes a robust technique for determining the alignment of 
two impurity moments in a gapped system. The details of the expected results
around magnetic impurities in the quasi-two-dimensional superconductor
NbSe$_2$ have been calculated. Energies and spatial structure of bonding
and antibonding states around parallel moments, and of localized atomic-like
states around antiparallel moments, indicate the two cases should be
distinguishable with nonmagnetic scanning tunneling spectroscopy. This 
technique should be broadly applicable to a wide range of correlated
electronic systems.

We would like to acknowledge the Office of Naval Research's
Grants Nos. N00014-96-1-1012 and N00014-99-1-0313. This research was 
supported in part by the National Science Foundation under Grant No.
PHY94-07194.

\begin{figure}
\caption[]{(color) 
Schematic of the potential for spin-up (left side, (A) and (C)) 
and spin-down
(right-side, (B) and (D)) 
quasiparticles in the presence of parallel impurity spins
(top row, (A) and (B)) and antiparallel impurity spins 
(bottom row (C) and (D)). For parallel
impurity spins there are two localized states of spin-up quasiparticles
which differ in energy, similar
to the bonding and antibonding states of a diatomic molecule. There are no
localized states of spin-down quasiparticles. For antiparallel impurity
spins there is one spin-down
quasiparticle localized state, as well as one of spin up, and 
the two are degenerate.}
\end{figure}

\begin{figure}
\caption[]{(color) (A) Energies of localized states as a function of impurity 
separation near parallel impurity
spins (red) and antiparallel impurity spins (black). Energy is in meV
and impurity separation in nearest-neighbor in-plane
lattice constants (3.47\AA). (B) Differential
density of states (DOS) for parallel impurity pairs (solid lines) and
antiparallel impurity pairs (dashed lines) for impurity separations
from one to five lattice spacings. (C) Same as (B), except for six to
ten lattice spacings.}
\end{figure}

\begin{figure}
\caption[]{(color) Spatial structure of the hole-like local 
density of states (LDOS) around
a single impurity in the surface layer of NbSe$_2$. Nearest-neighbor 
in-plane
separation on the triangular lattice is 3.47\AA. The units of the color scale
are eV$^{-1}$.}
\end{figure}

\begin{figure}
\caption[]{(color) LDOS around a parallel impurity
pair at (A) the energy corresponding to the bonding state ($-0.10$~meV), and
(B) the energy corresponding to the antibonding state ($-0.26$~meV). 
The impurities are at the same 
sites in each of (A-D), labeled $1$ and $2$ in (B).
The mirror plane between the impurities is indicated
by the red line in (B); there is no LDOS for the antibonding
state in (B) along this plane, while there is for the bonding state (A).
(C) LDOS around an antiparallel impurity pair 
at the energy of localized states ($-0.28$~meV). (D) 
spin-resolved LDOS at the same
energy as (C) showing the predominance of LDOS around the impurity on the
left. The units of the color scale are eV$^{-1}$.}
\end{figure}

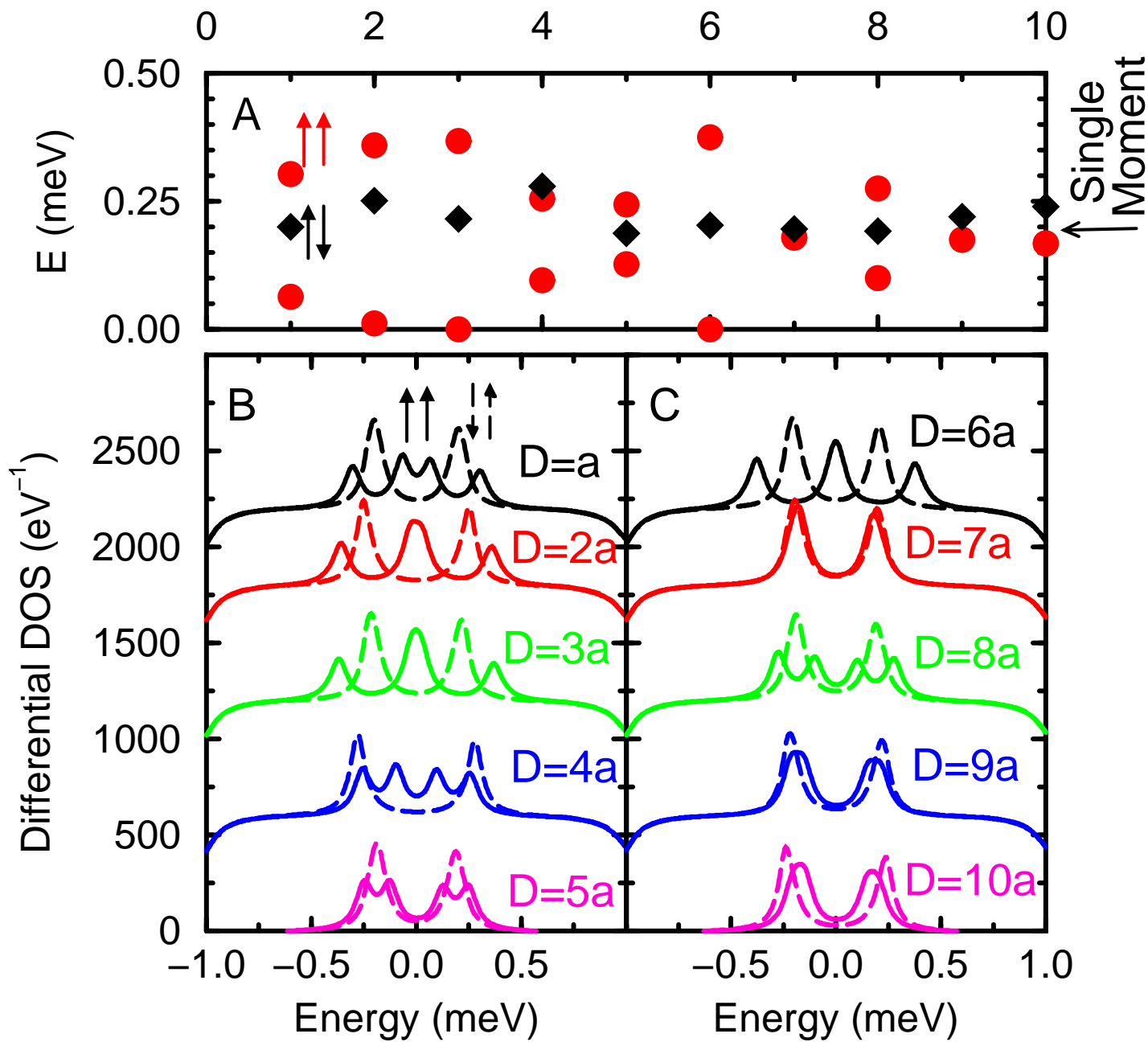
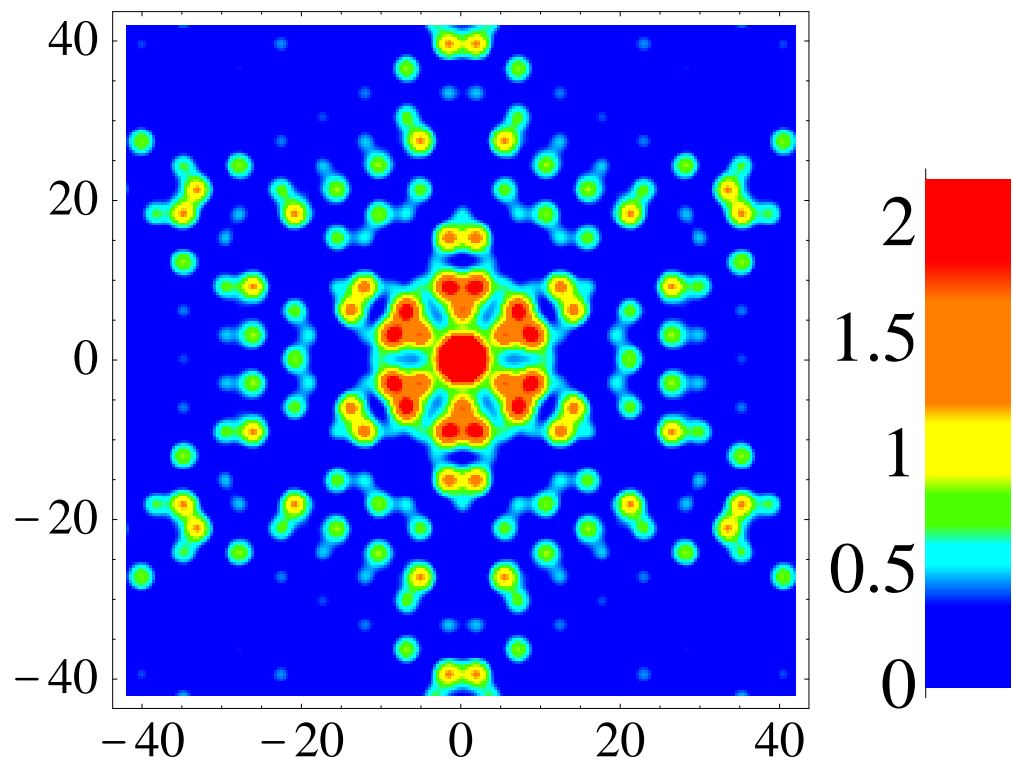
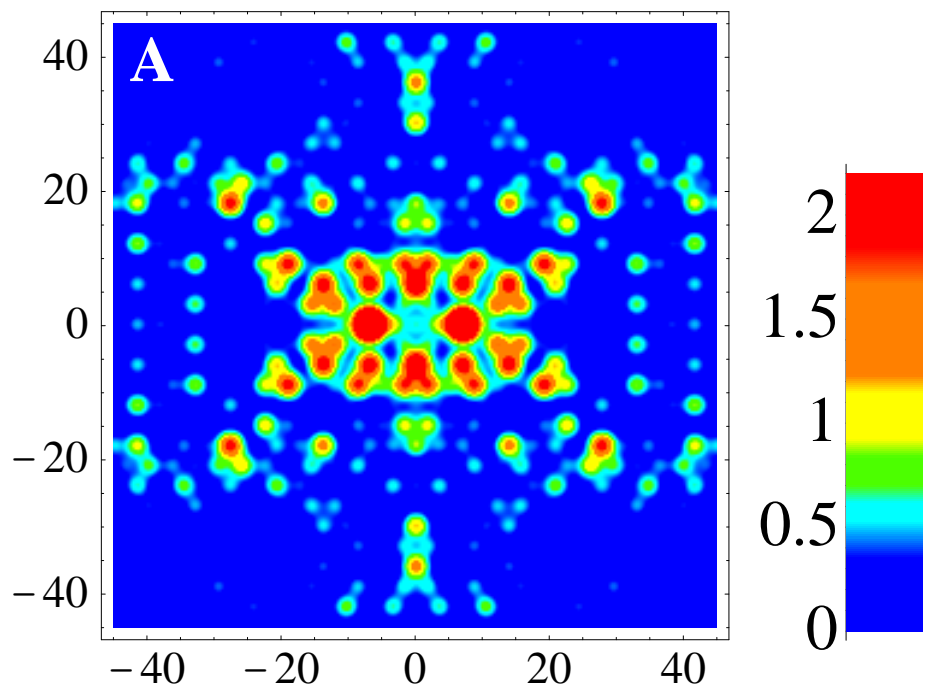
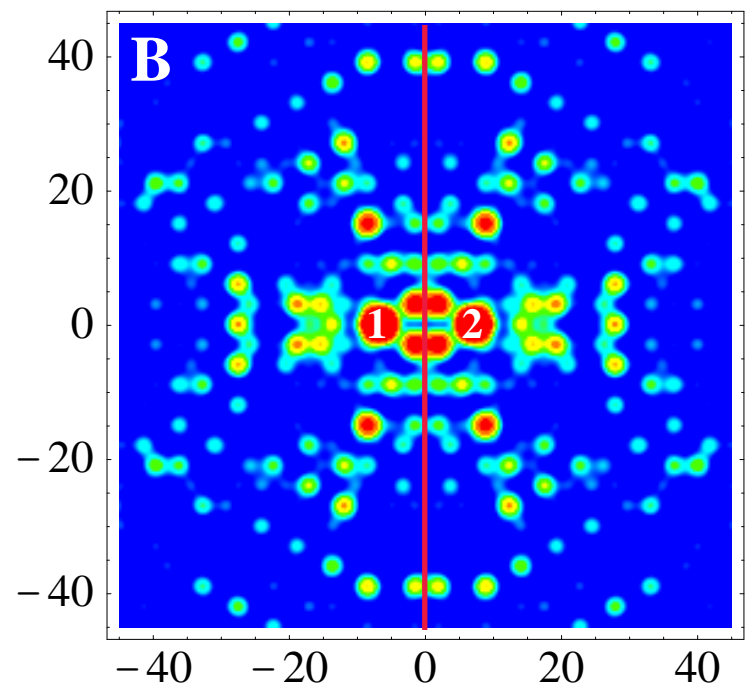
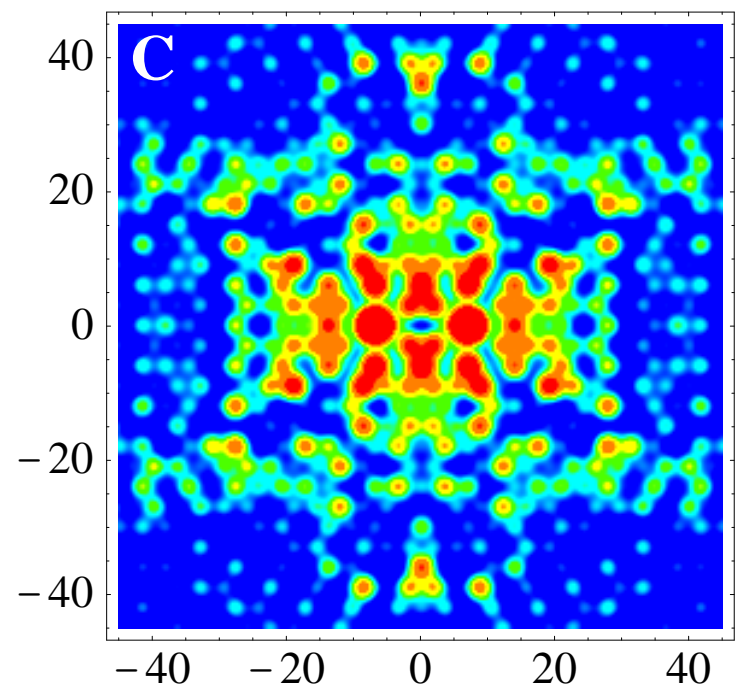
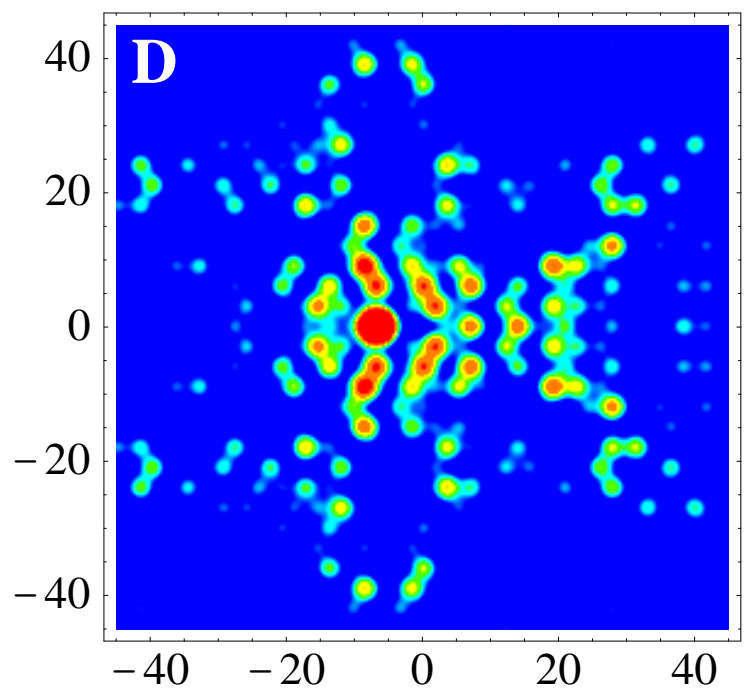
\end{document}